\documentclass[showpacs,preprintnumbers,amsmath,amssymb,nofootinbib]{revtex4}

\def\beq{\begin{equation}}
\def\eeq{\end{equation}}
\def\beqa{\begin{eqnarray}}
\def\eeqa{\end{eqnarray}}

\def\lsim{\mathrel{\raise.3ex\hbox{$<$\kern-.75em\lower1ex\hbox{$\sim$}}} }
\def\gsim{\mathrel{\raise.3ex\hbox{$>$\kern-.75em\lower1ex\hbox{$\sim$}}} }
\usepackage{color}
\usepackage{latexsym}
\usepackage{graphicx}
\usepackage{amssymb}
\usepackage{amsmath}

\begin{document}

\vspace*{.5in}
\title{Consequences of Approximate $S_{\scriptscriptstyle 3}$
Symmetry of the Neutrino Mass Matrix}

\vspace*{.5in}

\author{Chien-Yi Chen}
\author{L. Wolfenstein}
\affiliation{Department of Physics, Carnegie Mellon University,
Pittsburgh, Pennsylvania 15213}

\vspace*{1in}

\begin{abstract}
Assuming that the neutrino mass matrix is dominated by a term with
the permutation symmetry $S_{\scriptscriptstyle 3}$ it is possible
to explain neutrino data only if the masses are quasi-degenerate. A
sub-dominant term with an approximate $\mu -\tau$ symmetry leads to
an approximate tri-bimaximal form. Experimental consequences are
discussed.
\end{abstract}

\maketitle

\newpage

%\tableofcontents
%%%%%% Section 1 %%%%%%%%%%%%%%%%%%%%%%%%%%%%%%%%%%%%%%%%%%%%%%%%%%%%%
%\section{Introduction}
Early solar neutrino data suggested that one neutrino eigenstate
could be
\begin{equation}
S = \frac{1}{\sqrt{3}} \left( \nu_{\scriptscriptstyle e} +
\nu_{\scriptscriptstyle \mu} + \nu_{\scriptscriptstyle \tau} \right
)\; .\label{1}
\end{equation}
This led to the consideration of an $S_{\scriptscriptstyle 3}$
symmetry \cite{Wolfenstein:1978uw}. Today the MSW solution to the
solar neutrino problem has the higher-energy neutrinos emerging
from the sun in a state given to a good approximation by S. Here
we consider the possibility that the neutrino mass matrix is
dominated by a term with $S_{\scriptscriptstyle 3}$ symmetry
leading to S as an eigenstate. We then consider possible
perturbations that violate the symmetry.

Our assumption is that neutrino mass is due to new physics not
directly related to the origin of the masses of other particles. A
large number of papers \cite{Jora:2006dh} have presented detailed
models based on $S_{\scriptscriptstyle 3}$ symmetry. Here we do not
consider a model but simply try to relate possible symmetries of the
new physics to observations. The most general Majorana mass matrix
invariant under $S_{\scriptscriptstyle 3}$ is \beqa M_{0} &=& \left
(\begin{array}{ccccccc}
  A  & B & B \\
  B  & A & B \\
  B  & B & A
  \end{array} \right) \;. \label{m0} \eeqa
The eigenstates are necessarily \cite{Wolfenstein:1978uw} a singlet
given by S and a degenerate doublet D which can be chosen as
%\begin{equation}
%\begin{split}
\begin{align}
D_{\scriptscriptstyle a} &=  \frac {\nu_{\scriptscriptstyle \mu} -
\nu_{\scriptscriptstyle \tau}}{\sqrt{2}} \; , \tag{3a}\\
D_{\scriptscriptstyle b} &=  \sqrt{\frac{2}{3}}
\,\nu_{\scriptscriptstyle e} -
\sqrt{\frac{1}{6}}\,\left(\nu_{\scriptscriptstyle \mu}+
\nu_{\scriptscriptstyle \tau} \right )\; .\tag{3b} \label{3}
\end{align}
%\end{split}
%\end{equation}
\setcounter{equation}{3} The masses are
\begin{align} m_{\scriptscriptstyle S} &=  A+2B \; ,\tag{4a}
\\m_{\scriptscriptstyle D} &= A-B\; .\tag{4b}\label{4}
\end{align}
\setcounter{equation}{4} The eigenstates in Eq.(\ref{1}) and Eq.(3)
are those of the tri-bimaximal form of the mixing matrix discussed
in many papers \cite{Fritzsch:1998xs} as a fit to neutrino
oscillation data. However, in the fit the largest mass splitting is
that between $D_{\scriptscriptstyle a}$ and $D_{\scriptscriptstyle
b}$ responsible for the atmospheric neutrino oscillation with
smaller splitting between S and $D_{\scriptscriptstyle b}$
associated with the solar neutrino oscillation. We assume that the
breaking of the degeneracy is due to the perturbation that breaks
$S_{\scriptscriptstyle 3}$. In order that the $S_{\scriptscriptstyle
3}$ term dominate we require that all three masses start out
approximately equal by choosing
\begin{equation}
B = -2A+b\; ,\tag{5a}
\end{equation}
with $b\ll B$ so that
\begin{equation}
m_{\scriptscriptstyle D} \approx - m_{\scriptscriptstyle S} \approx
3A \; .\tag{5b}
\end{equation}
\setcounter{equation}{5}
The minus sign means that the state S has
the opposite CP eigenvalue from that of D. We have assumed here for
simplicity that A and B are real; otherwise D and S would have a
relative Majorana phase.
%p3
The sub-dominant mass matrix $M_{\scriptscriptstyle 1}$ that breaks
$S_{\scriptscriptstyle 3}$ has the result of raising the mass of one
D state above $m_{\scriptscriptstyle S}$ and leaving the mass of the
other slightly below $m_{\scriptscriptstyle S}$. These mass values
then correspond to what is called the "quasi-degenerate" case for
neutrino masses.

%%%%%%%%%%%%%%%%%%%%%%
We now assume that the perturbing matrix $M_{\scriptscriptstyle 1}$,
which is added to $M_{\scriptscriptstyle 0}$, breaks
$S_{\scriptscriptstyle 3}$ but retains a $S_{\scriptscriptstyle 2}$
symmetry between $\nu_{\scriptscriptstyle \mu}$ and
$\nu_{\scriptscriptstyle \tau}$. \beqa M_{1} &=& \left
(\begin{array}{ccccccc}
  e  & f & f \\
  f  & t & \epsilon \\
  f  & \epsilon & t
  \end{array} \right)  \label{m1} \eeqa
As a result of the symmetry $D_{\scriptscriptstyle a}$ remains an
eigenstate and the parameter known as $\theta_{\scriptscriptstyle
13}$ vanishes. In addition to providing the mass splitting between
$D_{\scriptscriptstyle a}$ and $D_{\scriptscriptstyle b}$,
$M_{\scriptscriptstyle 1}$ causes a small mixing of
$D_{\scriptscriptstyle b}$ with S. The parameters e and f can be
absorbed into A and B and so they are set to zero in what follows.
Matrices of the form $M_{\scriptscriptstyle 0}$ +
$M_{\scriptscriptstyle 1}$, with four parameters are discussed in
many papers \cite{Ma:2005pd}.

\begin{figure}
\begin{center}
%{\bf FIGURES}
%\bigskip
%\vspace{1cm}
\includegraphics[scale=1.0]{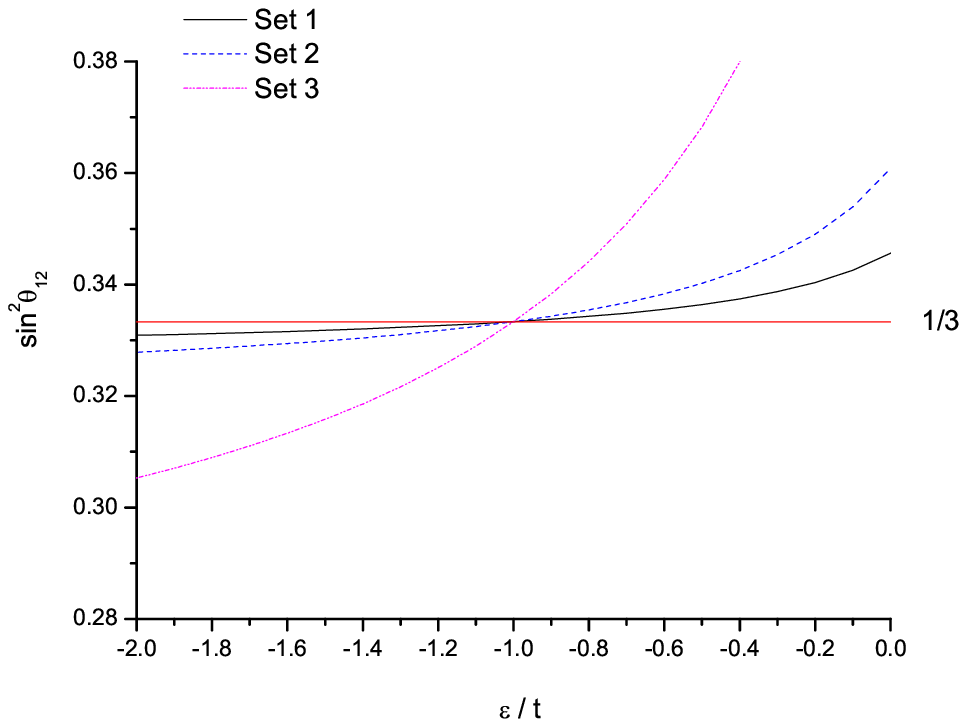}
\caption{The solar neutrino survival
${\sin}^{2}\,\theta_{\scriptscriptstyle 12}$ for the higher energy
neutrinos for the LMA-MSW solution as a function of
$\frac{\epsilon}{t}$.} \label{fig1}
\includegraphics[scale=1.0]{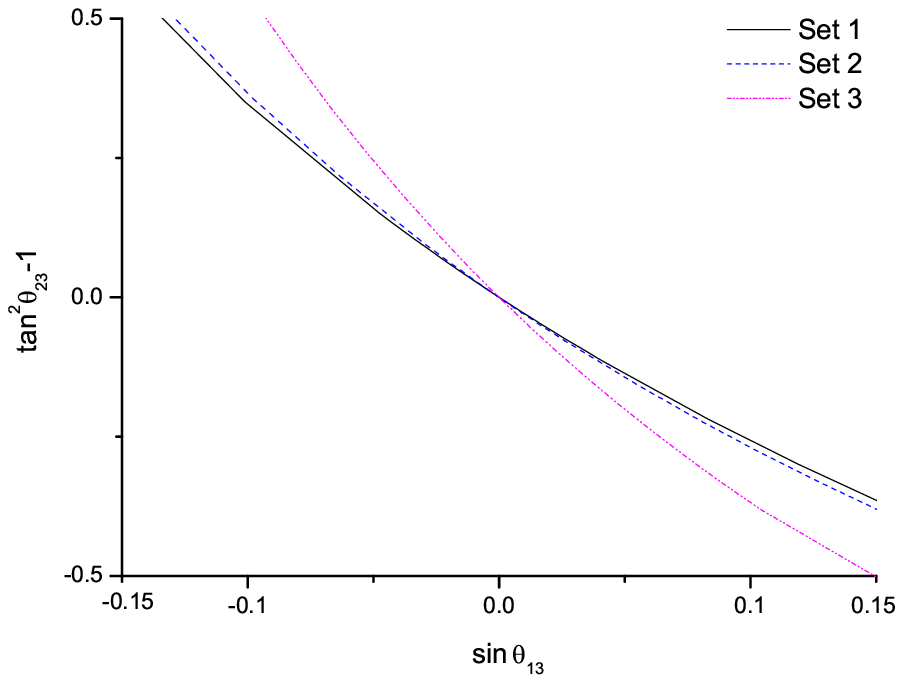}
\caption{Diagram with $\tan^{2}\,\theta_{\scriptscriptstyle 23}- 1$.
v.s $\sin \,\theta_{\scriptscriptstyle 13}.$ }\label{fig2}
\end{center}
\end{figure}

%%%%%%%% 6
We now identify the states which start out as
$(D_{\scriptscriptstyle a},S,D_{\scriptscriptstyle b})$ as
$(3,2,1)$. The mass $m_{\scriptscriptstyle 2}$ is understood to be
positive although $m_{\scriptscriptstyle S}$ is negative (assuming A
is positive). The mass differences are
\begin{align}
m_{\scriptscriptstyle 3}-m_{\scriptscriptstyle 1} &=  \frac {2}{3}
\; \left(t- 2\epsilon \right ) ,\tag{7a} \\
m_{\scriptscriptstyle 2}-m_{\scriptscriptstyle 1} &=  - \left(b + t
+ \epsilon \right ) \; .\tag{7b}\label{7}
\end{align}
\setcounter{equation}{7} The small value of $m_{\scriptscriptstyle
2}-m_{\scriptscriptstyle 1}$ required to fit the data involves the
fine-tuning of the value of b.
%p7
The resulting deviation of the factor $\frac{1}{\sqrt{3}}$ for
$\nu_{\scriptscriptstyle e}$ in $S$ is given approximately by
\beqa D = \frac{2}{3 \sqrt{3}} \left( \frac{t + \epsilon}{6 A}
\right )\; =\frac{k}{\sqrt{3}} \left( \frac{m_{\scriptscriptstyle
3}- m_{\scriptscriptstyle 1}}{2 m_{\scriptscriptstyle 2}} \right
)\; ,\label{8} \eeqa
\[
 k =\frac{t + \epsilon}{t -2 \epsilon} \; ,\]
\[
{\sin}^{2}\,\theta_{\scriptscriptstyle
12}=(\frac{1}{\sqrt{3}}+D)^{\,2}  \; .
\]
Since by our assumption of a quasi-degenerate neutrino mass
spectrum the mass ratio in Eq.(\ref{8}) is small so that D is
predicted to be small. To obtain the doublet mass splitting
without large parameters we choose $\frac{\epsilon}{t}$ to be
negative. As $\frac{\epsilon}{t}$ varies from 0 to a large
negative value k varies from 1 to $-\frac{1}{2}$; for
$\frac{\epsilon}{t}=-1$, D=0 and we obtain the tri-bimaximal form.
% p8
Choosing values for the mass-splittings fitted from oscillation
data \cite{GonzalezGarcia:2007ib}
\begin{equation}
\begin{split}
 m^{2}_{\scriptscriptstyle 3}-m^{2}_{\scriptscriptstyle 2}
&=2.6\times 10^{-3}eV^{2}\; , \\ m^{2}_{\scriptscriptstyle
2}-m^{2}_{\scriptscriptstyle 1}&=8\times 10^{-5}eV^{2}\; ,
\end{split}
\end{equation}
we give in table 1 three sets of mass values. The largest values
(like set 1) are limited by cosmology \cite{Hannestad:2006mi}
whereas the smallest values(like set 3) are limited by the
requirement that the magnitude of $M_{\scriptscriptstyle 1}$ is
smaller than $M_{\scriptscriptstyle 0}$. For each of these we show
in Fig. \ref{fig1} the solar neutrino survival
${\sin}^{2}\,\theta_{\scriptscriptstyle 12}$ for the higher energy
neutrinos for the LMA-MSW solution as a function of
$\frac{\epsilon}{t}$.
%p 8a
Note that the sign of the deviation from $\frac{1}{3}$ can be
either positive or negative. We have shown the case of the "normal
hierarchy" with ($m_{\scriptscriptstyle 3} - m_{\scriptscriptstyle
1}$) positive. In the case of the inverse hierarchy the curves are
flipped about the ${\sin}^{2}\,\theta_{\scriptscriptstyle
12}={\frac{1}{3}}$ axis. Assuming negligible Majorana phases the
mass that enters the double beta-decay formula is
\begin{equation}
m_{\scriptscriptstyle ee}=-{\sin}^{2}\,\theta_{\scriptscriptstyle
12}\; m_{\scriptscriptstyle 2}+
{\cos}^{2}\,\theta_{\scriptscriptstyle 12}\; m_{\scriptscriptstyle
1} \approx (1-2 \sin^{2}\,\theta_{\scriptscriptstyle 12})\;
m_{\scriptscriptstyle 2}\;, \label{mee}
\end{equation}
given the small difference between $m_{\scriptscriptstyle 2}$ and
$m_{\scriptscriptstyle 1}$.

% p9
We finally consider a possible small violation of $\mu-\tau$
symmetry by changing the 22 element in Eq.(\ref{m1}) to
$t+\frac{\delta}{2}$ and the 33 element to $t-\frac{\delta}{2}$. The
main effect is to mix $D_{\scriptscriptstyle a}$ and
$D_{\scriptscriptstyle b}$ or the states now labeled 3 and 1. There
is also a small mixing of 2 and 1 but this is suppressed by the
"mass difference" 6A. The important result is a non-zero value of
$\theta_{\scriptscriptstyle 13}$, the $\nu_{\scriptscriptstyle e}$
amplitude in state 3. Directly correlated with
$\theta_{\scriptscriptstyle 13}$ there is a deviation of
$\theta_{\scriptscriptstyle 23}$, the $\nu_{\scriptscriptstyle \mu}$
amplitude in state 3, from $\frac{\pi}{4}$.

Starting with the tri-bimaximal mixing, corresponding to the limit
$\frac{\epsilon}{t}=-1$, this correlation is given by \beq
{\tan}^{2}\,\theta_{\scriptscriptstyle 23} = 1-2 \sqrt{2} S + 4
S^{2}\; \label{10}\; ,\eeq
 \[ S =\sin\,\theta_{\scriptscriptstyle 13}\left(
\frac{1+2 \lambda}{1- \lambda} \right )\; ,\]
 \[\lambda =
\frac{m_{\scriptscriptstyle 3}- m_{\scriptscriptstyle 1}
}{m_{\scriptscriptstyle 2}+m_{\scriptscriptstyle 1}}\; ,\]
 to order
$S^{2}$. In Fig. 2, we show (${\tan}^{2}\,\theta_{\scriptscriptstyle
23} - 1$) as a function of $\sin\,\theta_{\scriptscriptstyle 13}$.
Different values of $\frac{\epsilon}{t}$ makes only small changes
since they are proportional to $\lambda D$. Given the small value of
$\theta_{\scriptscriptstyle 13}$ from experiment, there is less than
as 1 $\%$ contribution of $\theta_{\scriptscriptstyle 13}$ to
Eq.(\ref{mee}) for $m_{\scriptscriptstyle ee}$.

%%%%%%%%%%%%%%%%%%%%%%%%%%%%%%%%%%%%%%%%%%%%%%%%%%%%%%%%%%%%%%%%%%%%%%
%\section{Conclusion}
%%%%%%%%%%%%%%%%%%%%%%%%%%%%%%%%%%%%%%%%%%%%%%%%%%%%%%%%%%%%%%%%%%%%%%
In this paper we have looked at possible experimental signatures of
the assumption that the physics yielding the neutrino mass matrix
has a predominant $S_{\scriptscriptstyle 3}$ symmetry. We further
assume a sub-dominant term which breaks $S_{\scriptscriptstyle 3}$
but has an $S_{\scriptscriptstyle 2}$ $\mu -\tau$ symmetry. This
leads to

(1) The neutrino masses must be quasi-degenerate.

(2)\;\,$\theta_{\scriptscriptstyle 13}$, the
$\nu_{\scriptscriptstyle e}$ component in the atmospheric mixing,
vanishes and the mixing is maximal.

(3) The high-energy solar neutrino survival, governed by the LMA-MSW
solution, deviates only a little from $\frac{1}{3}$ as illustrated
in Fig. \ref{fig1}.

(4) In the absence of significant Majorana phases the mass
$m_{\scriptscriptstyle ee}$ governing double beta decay is
approximately equal to $\frac{m_{\scriptscriptstyle 2}}{3}$.

\noindent If we further allow a small term involving only
$\nu_{\scriptscriptstyle \mu}$ and $\nu_{\scriptscriptstyle \tau}$
that violates the $S_{\scriptscriptstyle 2}$ symmetry then there
is a non-zero $\theta_{\scriptscriptstyle 13}$. In this case the
atmospheric mixing angle is no longer maximum and its value is
directly correlated with $\theta_{\scriptscriptstyle 13}$ as shown
in Eq.(\ref{10}) and Fig. \ref{fig2}.

This work was supported in part by the Department of Energy with
Contact number DE-FG02-91ER40682.

%%%%%%%%%%%%%%%%%%%%%%%%%%%%%%%%%%%%%%%%%%%%%%%%%%%%%%%%%%
% The end
%%%%%%%%%%%%%%%%%%%%%%%%%%%%%%%%%%%%%%%%%%%%%%%%%%%%%%%%%%
%%%%%%%%%%%%%%%%%%%%%%%%%%%%%%%%%%%%%%%%%%%%%%%%%%%%%%%%%%

%%%%%%%%%%%%%%%%%%%%%%%%%%%%%%%%%%%%%%%%%%%%%%%%
%\end{document}
\newpage
%----------------------------------------------------------table 1
\begin{table}[bh]
\vspace*{.5in} \caption{\small \label{table1} Three sets of mass
values}
\bigskip
\centering
\begin{tabular}{cccc}
\;\; & $m_{\scriptscriptstyle 1}$ (eV) & $m_{\scriptscriptstyle 2}$ (eV)  & $m_{\scriptscriptstyle 3}$ (eV)  \\
\hline
1  & 0.1845 & 0.1847 & 0.1913 \\
\hline
2  & 0.1247 & 0.1250 & 0.1350 \\
\hline
3  & 0.0512 & 0.0520 & 0.0729 \\
\end{tabular}
\end{table}
%%%%%%%%%%%%%%%%%%%%% Figures %%%%%%%%%%%%%%%%%%%%%%
%%%%%%%%%%%%%%%%%%%%%%%%%%%%%%%%%%%%%%%%%%

\begin{thebibliography}{99}
%%%%%%%%%%%%%%%%%%%%%%%%%%%%%%%%%%%%% EXP
%1.0\cite{Wolfenstein:1978uw}
\bibitem{Wolfenstein:1978uw}
  L.~Wolfenstein,
  %``Oscillations Among Three Neutrino Types And CP Violation,''
  Phys.\ Rev.\  D {\bf 18}, 958 (1978).
  %%CITATION = PHRVA,D18,958;%%

%2.1\cite{Jora:2006dh}
\bibitem{Jora:2006dh}
  R.~Jora, S.~Nasri and J.~Schechter,
  %``An approach to permutation symmetry for the electroweak theory,''
  Int.\ J.\ Mod.\ Phys.\  A {\bf 21}, 5875 (2006)
  [arXiv:hep-ph/0605069] and references therein.
  %%CITATION = IMPAE,A21,5875;%%
%2.2\cite{Felix:2006pn}
%\bibitem{Felix:2006pn}
  O.~Felix, A.~Mondragon, M.~Mondragon and E.~Peinado,
  %``Neutrino masses and mixings in a minimal S(3)-invariant extension of the
  %standard model,''
  AIP Conf.\ Proc.\  {\bf 917}, 383 (2007)
  [Rev.\ Mex.\ Fis.\  {\bf S52N4}, 67 (2006)]
  [arXiv:hep-ph/0610061].
  %%CITATION = RMXFA,S52N4,67;%%
%2.3\cite{Chen:2004rr}
%\bibitem{Chen:2004rr}
  S.~L.~Chen, M.~Frigerio and E.~Ma,
  %``Large neutrino mixing and normal mass hierarchy: A discrete
  %understanding,''
  Phys.\ Rev.\  D {\bf 70}, 073008 (2004)
  [Erratum-ibid.\  D {\bf 70}, 079905 (2004)]
  [arXiv:hep-ph/0404084].
  %%CITATION = PHRVA,D70,073008;%%
%2.4\cite{Grimus:2005mu}
%\bibitem{Grimus:2005mu}
  W.~Grimus and L.~Lavoura,
  %``S(3) x Z(2) model for neutrino mass matrices,''
  JHEP {\bf 0508}, 013 (2005)
  [arXiv:hep-ph/0504153].
  %%CITATION = JHEPA,0508,013;%%
%2.5\cite{Mohapatra:2006pu}
%\bibitem{Mohapatra:2006pu}
  R.~N.~Mohapatra, S.~Nasri and H.~B.~Yu,
  %``S(3) symmetry and tri-bimaximal mixing,''
  Phys.\ Lett.\  B {\bf 639}, 318 (2006)
  [arXiv:hep-ph/0605020].
  %%CITATION = PHLTA,B639,318;%%
%2.6\cite{Caravaglios:2005gw}
%\bibitem{Caravaglios:2005gw}
  F.~Caravaglios and S.~Morisi,
  %``Neutrino masses and mixings with an S(3) family permutation symmetry,''
  arXiv:hep-ph/0503234.
  %%CITATION = HEP-PH/0503234;%%
%2.7\cite{Fritzsch:1999im}
%\bibitem{Fritzsch:1999im}
  H.~Fritzsch and Z.~z.~Xing,
  %``Maximal neutrino mixing and maximal CP violation,''
  Phys.\ Rev.\  D {\bf 61}, 073016 (2000)
  [arXiv:hep-ph/9909304].
  %%CITATION = PHRVA,D61,073016;%%
%2.8\cite{Picariello:2006sp}
%\bibitem{Picariello:2006sp}
  M.~Picariello,
  %``Neutrino CP violating parameters from nontrivial quark-lepton correlation:
  %A S(3) x GUT model,''
  arXiv:hep-ph/0611189.
  %%CITATION = HEP-PH/0611189;%%

%3.1\cite{Fritzsch:1998xs}
\bibitem{Fritzsch:1998xs}
  H.~Fritzsch and Z.~z.~Xing,
  %``Large leptonic flavor mixing and the mass spectrum of leptons,''
  Phys.\ Lett.\  B {\bf 440}, 313 (1998)
  [arXiv:hep-ph/9808272].
  %%CITATION = PHLTA,B440,313;%%
%3.2\cite{He:2003rm}
%\bibitem{He:2003rm}
  X.~G.~He and A.~Zee,
  %``Some simple mixing and mass matrices for neutrinos,''
  Phys.\ Lett.\  B {\bf 560}, 87 (2003)
  [arXiv:hep-ph/0301092].
  %%CITATION = PHLTA,B560,87;%%
%3.3\cite{Harrison:1999cf}
%\bibitem{Harrison:1999cf}
  P.~F.~Harrison, D.~H.~Perkins and W.~G.~Scott,
  %``A redetermination of the neutrino mass-squared difference in  tri-maximal
  %mixing with terrestrial matter effects,''
  Phys.\ Lett.\  B {\bf 458}, 79 (1999)
  [arXiv:hep-ph/9904297].
  %%CITATION = PHLTA,B458,79;%%
%3.4\cite{Harrison:2002er}
%\bibitem{Harrison:2002er}
  P.~F.~Harrison, D.~H.~Perkins and W.~G.~Scott,
  %``Tri-bimaximal mixing and the neutrino oscillation data,''
  Phys.\ Lett.\  B {\bf 530}, 167 (2002)
  [arXiv:hep-ph/0202074].
  %%CITATION = PHLTA,B530,167;%%
%3.5\cite{Harrison:2003aw}
%\bibitem{Harrison:2003aw}
  P.~F.~Harrison and W.~G.~Scott,
  %``Permutation symmetry, tri-bimaximal neutrino mixing and the S3 group
  %characters,''
  Phys.\ Lett.\  B {\bf 557}, 76 (2003)
  [arXiv:hep-ph/0302025].
  %%CITATION = PHLTA,B557,76;%%

%4.1\cite{Ma:2005pd}
\bibitem{Ma:2005pd}
  E.~Ma,
  %``Neutrino mass matrix from S(4) symmetry,''
  Phys.\ Lett.\  B {\bf 632}, 352 (2006)
  [arXiv:hep-ph/0508231].
  %%CITATION = PHLTA,B632,352;%%
%4.2%\cite{Altarelli:2007cd}
%\bibitem{Altarelli:2007cd}
  G.~Altarelli,
  %``Models of neutrino masses and mixings: A progress report,''
  arXiv:0705.0860 [hep-ph].
  %%CITATION = ARXIV:0705.0860;%%
%4.3\cite{Grimus:2003kq}
%\bibitem{Grimus:2003kq}
  W.~Grimus and L.~Lavoura,
  %``A discrete symmetry group for maximal atmospheric neutrino mixing,''
  Phys.\ Lett.\  B {\bf 572}, 189 (2003)
  [arXiv:hep-ph/0305046].
  %%CITATION = PHLTA,B572,189;%%

%5.0%\cite{GonzalezGarcia:2007ib}
\bibitem{GonzalezGarcia:2007ib}
  M.~C.~Gonzalez-Garcia and M.~Maltoni,
  %``Phenomenology with Massive Neutrinos,''
  arXiv:0704.1800 [hep-ph].
  %%CITATION = ARXIV:0704.1800;%%

%6.1\cite{Hannestad:2006mi}
\bibitem{Hannestad:2006mi}
  S.~Hannestad and G.~G.~Raffelt,
  %``Neutrino masses and cosmic radiation density: Combined analysis,''
  JCAP {\bf 0611}, 016 (2006)
  [arXiv:astro-ph/0607101].
  %CITATION = JCAPA,0611,016;%%
%6.2\cite{Fukugita:2006rm}
%\bibitem{Fukugita:2006rm}
  M.~Fukugita, K.~Ichikawa, M.~Kawasaki and O.~Lahav,
  %``Limit on the neutrino mass from the WMAP three year data,''
  Phys.\ Rev.\  D {\bf 74}, 027302 (2006)
  [arXiv:astro-ph/0605362].
  %%CITATION = PHRVA,D74,027302;%%
%6.3\cite{Seljak:2006bg}
%\bibitem{Seljak:2006bg}
  U.~Seljak, A.~Slosar and P.~McDonald,
  %``Cosmological parameters from combining the Lyman-alpha forest with CMB,
  %galaxy clustering and SN constraints,''
  JCAP {\bf 0610}, 014 (2006)
  [arXiv:astro-ph/0604335].
  %%CITATION = JCAPA,0610,014;%%
%6.4\cite{Goobar:2006xz}
%\bibitem{Goobar:2006xz}
  A.~Goobar, S.~Hannestad, E.~Mortsell and H.~Tu,
  %``A new bound on the neutrino mass from the SDSS baryon acoustic peak,''
  JCAP {\bf 0606}, 019 (2006)
  [arXiv:astro-ph/0602155].
  %%CITATION = JCAPA,0606,019;%%

%%%%%%%%%%%%%%%%%%%%%%%%%%%%%%%%%%% Recent research develope
%\cite{Choudhury:2006sq}
\end{thebibliography}
\end{document}